\documentclass[letterpaper, 10 pt, conference]{ieeeconf}

\IEEEoverridecommandlockouts                              

\usepackage{graphicx}
\usepackage{amssymb}
\usepackage[font=footnotesize]{caption}
\usepackage{subcaption}

\title{ \LARGE \bf On the use of higher-order tensors to model muscle synergies }

\author{Ahmed Ebied,\textit{ student member, IEEE},
		Loukianos Spyrou,
		Eli Kinney-Lang,\textit{ student member, IEEE}, 
\\
		\& Javier Escudero,\textit{ member, IEEE}
\thanks{ A. Ebied, L. Spyrou, E. Kinney-Lang, and J. Escudero are with the Institute for Digital Communications, School of Engineering, University of Edinburgh, Edinburgh EH9 3FB, United Kingdom	  
(Corresponding author; email: {\tt\small ahmed.ebied@ed.ac.uk})}%
\thanks{This study was partially supported by the Engineering and Physical Sciences Research Council (EPSRC) via the research project EP/N014421/1.}	
		}

\begin{document}

\maketitle
\thispagestyle{empty}
\pagestyle{empty}

\begin{abstract}
The muscle synergy concept provides the best framework to understand motor control and it has been recently utilised in many applications such as prosthesis control. The current muscle synergy model relies on decomposing multi-channel surface Electromyography (EMG) signals into a synergy matrix (spatial mode) and its weighting function (temporal mode). This is done using several matrix factorisation techniques, with Non-negative matrix factorisation (NMF) being the most prominent method. Here, we introduce a $4^{th}$-order tensor muscle synergy model that extends the current state of the art by taking spectral information and repetitions (movements) into account. This adds more depth to the model and provides more synergistic information. In particular, we illustrate a proof-of-concept study where the Tucker3 tensor decomposition model was applied to a subset of wrist movements from the Ninapro database. The results showed the potential of Tucker3 tensor factorisation in finding patterns of muscle synergies with information about the movements and highlights the differences between the current and proposed model.
\end{abstract}

\section{Introduction}

The question of how the central nervous system (CNS) controls body movements has been discussed for over a century with no conclusive answer. This is due to the high level of complexity and dimensionality of the motor control associated with any movement due to the multiple muscles and possible degrees of freedom (DoFs) \cite{DAvella2015}. So far, the muscle synergy concept \cite{Tresch1999,Saltiel2001} provides the best explanation for the motor control process. According to this concept, the CNS does not control each muscle directly to generate a motor output. Rather, the CNS controls synergies where each synergy activates a group of muscles and the combination of those synergies is responsible for the motor output. This approach maps the sensory input onto motor output through an internal model. This model is considered to be in a lower dimensional subspace and synergies are the basis of that subspace.

Two mathematical muscle synergy models have been proposed: time-invariant \cite{Tresch1999} and time-varying \cite{DAvella2002}. The time-invariant model defines the muscle activity as synchronised synergies weighted by time-varying coefficients. The time-varying model describes the muscle activity as a combination of asynchronous synergies compromised by a collection of scaled and shifted waveforms. Although the time-varying model provides more parsimonious representation for the muscle compared to the time-invariant model, some studies show evidence that the muscle synergies are synchronised in time \cite{Kargo2008}. Therefore, most recent muscle synergies studies rely on the time-invariant model. 

The extraction of muscle synergies can be considered as a blind source separation problem where the recorded muscle activity is used to estimate the unknown underlying synergies and their weighting functions. This is traditionally approached by applying matrix factorisation on the multichannel EMG recordings which are represented as a matrix with spatial (channels) and temporal (samples) modes.

Several matrix factorisation techniques, such as principal component analysis (PCA), independent component analysis (ICA) \cite{Hyvarinen2000} and non-negative matrix factorisation (NMF) \cite{Lee1999}, have been applied to identify the muscle synergies \cite{Tresch2006}. In recent years, the muscle synergy concept has been used in rehabilitation research including prosthesis control. This is carried out by applying matrix factorisation techniques on multichannel EMG recordings for specific movements. Synergistic information is extracted by NMF \cite{Jiang2014b,Choi2011}, PCA \cite{Matrone2012} or ICA \cite{Rasool2016} and used as training/learning data to either classify testing tasks \cite{Choi2011,Rasool2016} or estimate proportional control signals for myoelectric control system \cite{Jiang2014b,Ma2015a}.

The key limitation of these approaches is the number of electrodes and synergies needed to approximate well the muscles pattern. Muscle synergy based myoelectric control systems usually require a high number of electrodes to collect enough spatial information to estimate synergies explaining most of the variance in the data. Moreover, there was a significant degradation in performance with the increase of task-dimensionality in \cite{Jiang2014b,Ma2015a}  due to the crosstalk effect between channels. The effect of synergies/channels number on the myoelectric control have been discussed in \cite{DeRugy2013} which shows that the number of synergies increases with task-space dimensionality to a point where no dimensionality reduction occurs in a 3-dimension task space.

We hypothesise that by adding more depth and domains to extract information from the muscle synergy model, it would be possible to improve the extracted synergistic information. This would be carried out by taking into account aspects other than the spatial profile of the synergies. An obvious candidate to consider is the spectral profile, which could provide a model that is robust to frequency changes due to fatigue and channel cross-talk. Consequently, here we expand the current muscle synergy model into a higher order one where synergies are estimated through tensor decomposition rather than matrix ($2^{nd}$-order tensor) factorisation. Although tensor factorisation had been applied on several biomedical signals studies such as EEG \cite{Escudero2015}, it has been investigated on EMG in only one study \cite{Xie2013a}, in which it was used for feature extraction from a 2-channels EMG signal for classification.

Here, a $4^{th}$-order tensor muscle synergy model is introduced, which expands the current model into 4 modes by adding spectral and repetitions modes into the spatial and temporal modes. In addition, a comparison of the models was done to classify the wrist movements. This study presents a proof-of-concept for the use of tensor factorisations for the extraction of muscle synergies.  

\section{Methodology}

We begin by reporting the publicly available dataset that has been analysed. Then the $2^{nd}$-order muscle synergy model is presented to compare it with the higher order model, which is discussed in the next two sections. The tensorisation of the EMG data and the Tucker3 model \cite{Tucker1966} that have been applied are described followed by the classification process.

\subsection{Dataset}

The Ninapro first dataset \cite{Atzori2012} is used which consists of recordings for 62 wrist, hand and finger movements. The analysed time series is a Root-Mean-Square (RMS) rectified version of the 10 channels raw surface EMG signal sampled at 100 Hz. The wrist motion is investigated. Consequently, 6 movements have been selected from the dataset representing the 3 main DoFs of the wrist: the wrist flexion and extension (DoF1), the wrist radial and ulnar deviation (DoF2) and the wrist  supination and pronation (DoF3). Each movement has 10 repetitions for each of the 27 healthy subjects.

\subsection{ $2^{nd}$-order model}

According to the time-invariant model, the muscle activity of the $j^{th}$ channel $\textbf{x}_{j}(n)$ is considered as a combination of $r$ synchronous synergies $\textbf{s}$ scaled by a set of time varying coefficients $\textbf{w}(n)$ \cite{Tresch1999} as shown in  (\ref{eq_1}).
\begin{equation}\label{eq_1}
\textbf{x}_{j}(n)=\sum_{i=1}^{i=r}s_{ij}\textbf{w}_i(n)
\end{equation}
Hence, the multichannel EMG signals represented as a matrix $\mathbf{X}$ with dimensions (\textit{m} channels $\times$ \textit{n} samples). According to this model, the multichannel EMG recordings are factorised into two lower rank matrices. The synergy matrix $\mathbf{S}$, which holds the channel (spatial) profile and the weighting matrix $\mathbf{W}$ with the temporal profile as shown in (\ref{eq_2}).
\begin{equation}\label{eq_2}
\mathbf{X}_{(m\times n)} = \mathbf{S}_{(m\times r)} \times\textbf{W}_{(r\times n)}
\end{equation}

The most widely used matrix factorisation techniques for synergy estimation is the NMF \cite{Lee1999}. The non-negativity constraints make it the most appropriate factorisation method due to the additive and non-negative nature of synergies \cite{Tresch2006}. 

\subsection{Data Tensorisation}

Tensors are a higher-order generalisation of vectors ($1^{st}$-order) and matrices ($2^{nd}$-order). They will be denoted as $\underline{\mathbf{X}}\in \mathbb{R}^{I_{1}\times I_{2} \times ... I_{n}} $ where $n \ge 3$. Therefore, to move beyond the  $2^{nd}$-order model, the first step is to create a higher-order synergy model by preparing the data in higher order tensor form. In this study, we  take spectral information into account for synergy identification. We hypothesise that synergies have distinct spectral components since motor unit action potential firing rate relies on the muscle's force modulation \cite{Kukulka1981}. Therefore, a time-frequency analysis technique is used to estimate the spectral components. Wavelet analysis is applied to each EMG channel activity using the Log-normal wavelet as a mother wavelet. Since it has a logarithmic frequency resolution, it provides increased frequency resolution compared to linear wavelets \cite{Iatsenko2015}. The 5-seconds epochs decomposed into 282 frequency-bins between 0 to 50 Hz. 

This converts the multichannel EMG epoch $\mathbf{X}\in\mathbb{R}^{I_{1}\times I_{2}}$ into $3^{rd}$-order tensor $\underline{\mathbf{X}}\in \mathbb{R}^{I_{1}\times I_{2} \times I_{3}} $ where each slice of this tensor is the wavelet transform for the respective channel. Then, by concatenating these $3^{rd}$-order tensors, a $4^{th}$-order tensor $\underline{\mathbf{X}}\in \mathbb{R}^{I_{1}\times I_{2} \times I_{3} \times I_{4}}$ is constructed. The additional (repetition) mode identifies the movement itself. Here, a $4^{th}$-order tensor is created for each DoF of the wrist's 3 main DoFs. The repetition mode consist of 12 repetitions divided equally between the positive and negative movements of the DoF. For example, 6 EMG epochs for each of wrist flexion and extension are tensorised into a $3^{rd}$-order tensor to add spectral mode then concatenated to form a $4^{th}$-order tensor for DoF1 with dimensions (10 channels$\times$500 samples$\times$282 frequency bins$\times$12 repetitions).

\subsection{Higher-order model}

 Tensors can be factorised into its main components like matrices. Naively, the easiest way to do this is by unfolding the tensor into a matrix and apply a matrix factorisation technique. However, this approach would discard any information from the mutual interactions between the higher dimensions \cite{Cichocki2014} and would not utilise the power of higher-order tensors.

Several higher-order tensor decomposition models have been introduced \cite{Kolda2008c}. The Tucker3 model is one of the most prominent ones. In the Tucker3 model \cite{Tucker1966}, the tensor is decomposed into a core tensor multiplied (transformed) by a matrix along each mode (dimension). For example, a $3^{rd}$-order tensor $\underline{\mathbf{X}}\in \mathbb{R}^{I_{1}\times I_{2} \times I_{3}} $  can be decomposed into smaller core tensor $\underline{\mathbf{G}}\in \mathbb{R}^{J_{1}\times J_{2} \times J_{3}} $ and three factor matrices of $ \textbf{B}^{(1)}\in \mathbb{R}^{I_{1}\times J_{1}}$, $ \textbf{B}^{(2)}\in \mathbb{R}^{I_{2}\times J_{2}}$ and $ \textbf{B}^{(3)}\in \mathbb{R}^{I_{3}\times J_{3}}$. In general, any tensor $\underline{\mathbf{X}}\in \mathbb{R}^{I_{1}\times I_{2} \times .... I_{N}}$ can be expressed as 

\begin{equation}\label{eq_3}
\underline{\mathbf{X}} \approx \underline{\mathbf{G}} \times {} _{1}\textbf{B}^{(1)} \times {} _{2}\textbf{B}^{(2)} .... \times {} _{n}\textbf{B}^{(n)} 
\end{equation}

Unlike other tensor decomposition models, the Tucker model is flexible in determining the number of component for each mode. However, the solution for Tucker3 is not unique. Therefore, adding constraints  to the model reduces the possibility of numerical degeneracy \cite{Kolda2008c}. In addition, non-negative factors are more appropriate with the  physiological significance similar to NMF. 

\begin{figure*}[t]
	\centering
	\includegraphics[width=1\linewidth]{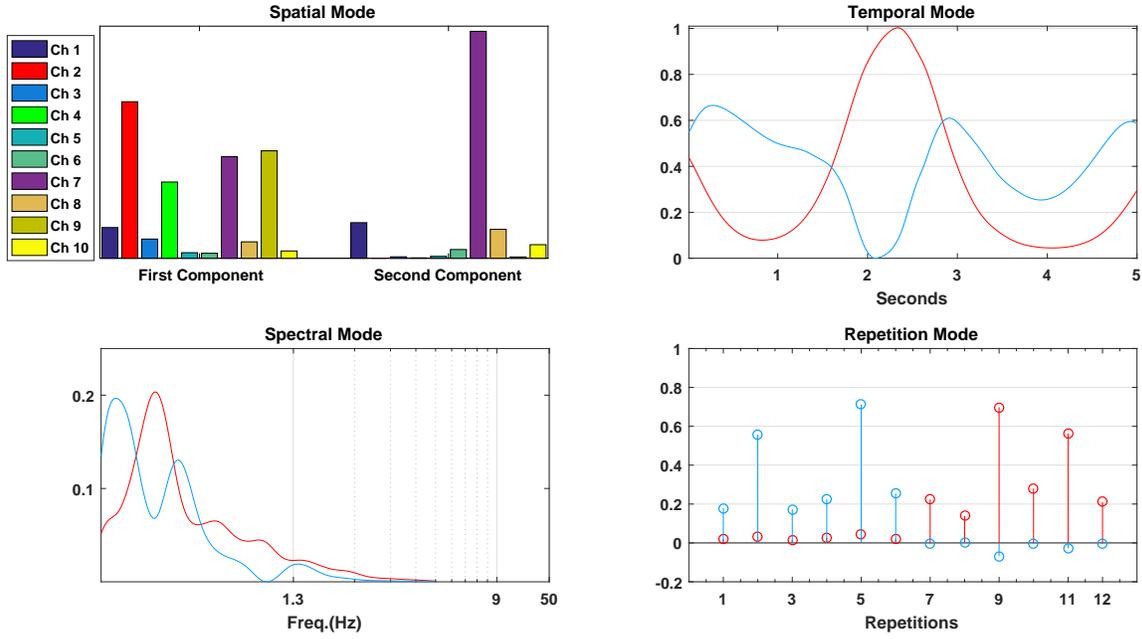}
	\caption{ The factors extracted via a Tucker3 decomposition of the $4^{th}$-order tensor for DoF1 (subject 4) assuming two components per mode. The repetition mode components showed the difference between the two movements where the blue component has a higher values for the wrist flexion repetitions (1-6) while the red was higher in the wrist extension repetitions (6-12). }
	\label{fig:4thordertensordecomp}
\end{figure*}

\subsection{Classification}

For each subject, the dataset is divided into training and testing sets where 60\% of the data have been assigned to training and 40\% are testing. The data is tensorised to form separate training and testing $4^{th}$-order tensors with modes channels$\times$samples$\times$frequencies$\times$repetitions. For a single DoF, it consists of 2 movements (positive and negative) as the repetitions of both movements are combined in one $4^th$-order tensor.

The Tucker3 model is applied on the training $4^{th}$-order tensor. For simplicity,  2 components were chosen for each mode, since preliminary results showed subtle differences with the change in number of components. The extracted core tensor and the components of all modes except "repetitions" are used to estimate the "repetitions" mode for the testing data. This is done by projecting the testing tensor onto the fixed training components (core tensor and first three modes).

The values of training repetition's mode components are used to train a \textit{k}-nearest neighbours (\textit{k}-NN) classifier (\textit{k}=3). While the testing "repetitions" mode components values are used as predictor to classify each repetition into either the positive or the negative movement of the DoF. 

As a benchmark, the same classifier has been trained by the synergy matrices  extracted from the training dataset using NMF. The number of synergies extracted was 1 for each movement (2 for each DoF) as in \cite{Jiang2014b} and to have the same number of factors (components) as the higher-order tensor classification method. NMF is applied on the testing dataset to estimate the synergy matrices which have used as a predictor to the movement. 

This has been carried out for the 27 healthy subjects. The 3 DoFs of the wrist are investigated where each DoF consists of two movements (positive and negative) and the classification accuracy is calculated for each DoF.

\begin{figure}[ht]
	\centering
	\begin{subfigure}[b]{\linewidth}
		\includegraphics[width=\textwidth]{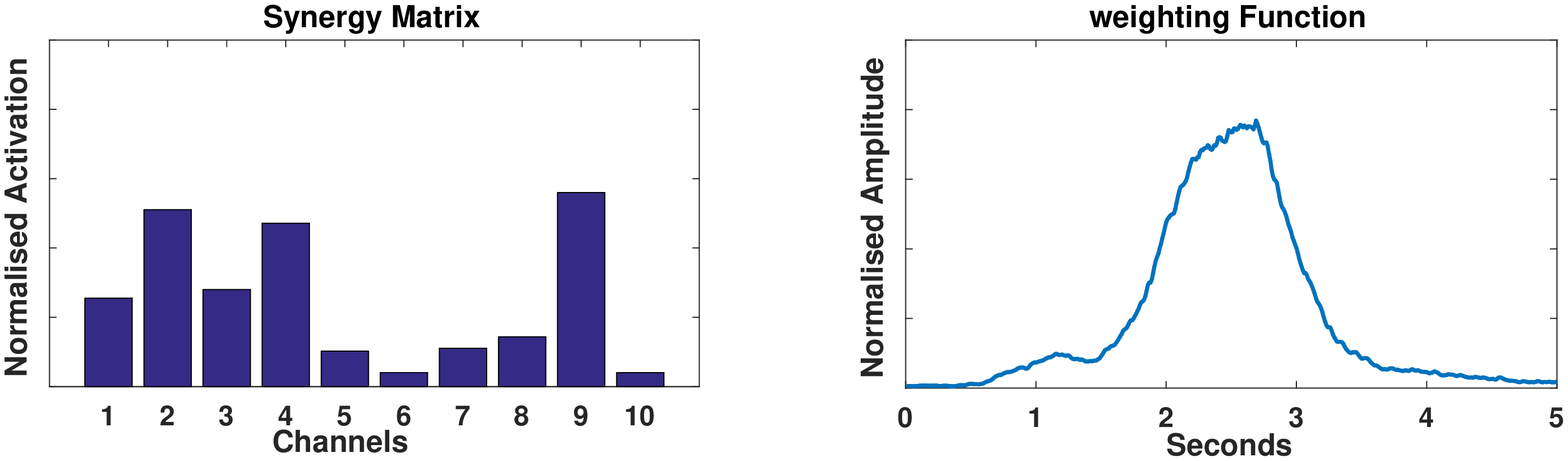}
		\caption{Wrist flexion (Positive DoF) NMF}
		\label{fig:posdofnmf}
	\end{subfigure}

	\hfill
	
	\begin{subfigure}[b]{\linewidth}
		\includegraphics[width=\textwidth]{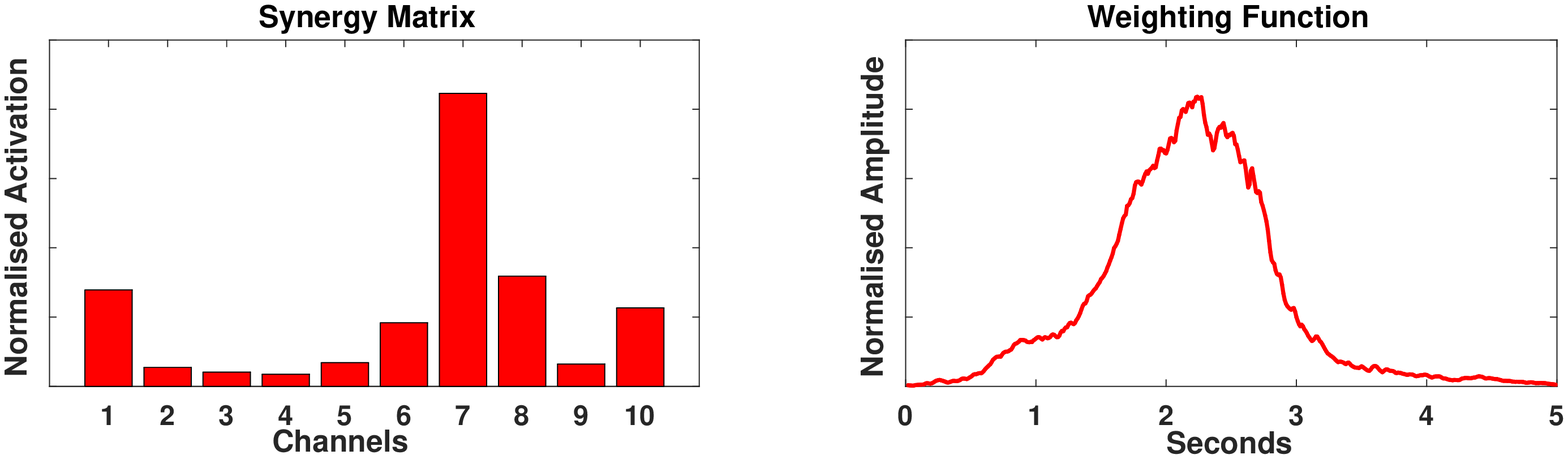}
		\caption{Wrist extension (Negative DoF) NMF}
		\label{fig:negdofnmf}
	\end{subfigure}
	\caption{The average synergy matrices (spatial profile) and weighting functions (temporal profile) estimated by one component NMF for DoF1 training dataset (subject 4).}
	\label{figure_NMF}
\end{figure}

\section{Results and Discussion}

Using the Tucker3 method, the tensors are decomposed into a smaller core tensor and 2-factor matrices for each mode (spatial, temporal, spectral and repetitions) as shown in Fig. \ref{fig:4thordertensordecomp}. The spatial and temporal modes (synergy matrix and weighting function) estimated via NMF for the same training data for each movement and the average across repetitions are shown in Fig. \ref{figure_NMF}.

There are differences and similarities between the estimated synergies from both models. If we compare the common synergy modes (spatial and temporal), we notice two points. Firstly, there is a similarity between spatial components from Tucker3 model and the synergy matrices (spatial components) of the NMF model for each separate movement. Although the Tucker3 components do not have to be directly associated with the movements, the spatial mode estimated the 2 components each is strongly linked to one movement. Secondly, unlike the spatial mode, the 2 components of temporal mode in the tensor model are not linked directly to movements. This is due to the similar weighting functions (temporal components)  for each movement. Thus, the first temporal component (red) in the Tucker3 model represents both weighting functions while the second component (blue) expresses other activities in this time window.  The same concept could be applied on the spectral mode where the two components represent the lower and higher frequency elements in the data. This suggests that the tensor decompositions are better able to reveal patterns in the EMG envelops than matrix factorisations.

The tensor approach exhibited a slight improvement in classification accuracy over NMF as shown in Table \ref{table_result}. This suggests a greater utility of higher-order tensor models taking also into account that such models can provide more descriptive information.

The current muscle synergy model relies only on the spatial information from the synergy matrices to deduce motor control. This approach is vulnerable to many factors such as electrode repositioning, sweat and fatigue \cite{DeRugy2013}. Therefore, other factors and variables should be taken into account. The higher-order tensor muscle synergy models provide the opportunity to alleviate the effect of those factors by incorporating a more complex description of the data and its dependencies. 

\begin{table}[h]
	\renewcommand{\arraystretch}{1.3}
	\caption{Average classification error rate across the 27 subjects.}
	\label{table_result}
  	\centering
\begin{tabular}{cccc}
	\hline 
	& DoF1 & DoF2 & DoF3 \\ 
	& (wrist flexion & (wrist radial and & (wrist supination\\
	& and extension) & ulnar deviation) &  and pronation)\\
	\hline 
	Tucker3 & 0\% & 0\% &   0\%\\ 
	NMF & 0.463\% & 0.463 \% & 2.315\% \\ 
	\hline 
\end{tabular} 
\end{table}
\section{Conclusion}
These results are limited by the off-line data and the small number of movements and DoFs. However, this study provides a proof-of-concept for higher-order tensor muscle synergy models. The complexity of these models could increase the computational cost but the additional modes offer a new range of possibilities to incorporate more information. These results suggest that tensor factorisation models can be a useful tool for muscle synergies extraction. In addition, it is encouraging to explore the potential of higher-order tensor muscle synergy model in future work. 

\bibliographystyle{ieeetr}
\bibliography{ConfMendely}

\end{document}